\documentclass[twocolumn]{aastex62}

\usepackage[utf8]{inputenc}
\DeclareUnicodeCharacter{2212}{-}
\usepackage{graphicx}
\graphicspath{{./figures/}{./}}
\usepackage{amsmath}
\usepackage{enumitem}
\setenumerate{itemsep=0mm}
\hypersetup{urlcolor={blue}}

\usepackage{soul} 
\usepackage{amsmath}
\usepackage{amssymb}
\usepackage{xspace}
\usepackage{xifthen}
\usepackage{eso-pic}

\newcommand{\bliss}{{BLISS}\xspace}
\newcommand{\Gaia}{{\it Gaia}\xspace}
\newcommand{\iauname}{{\bliss~J0321$+$0438}\xspace}
\newcommand{\juicy}{{\iauname}\xspace (\bliss 1)\xspace}

\definecolor{forestgreen}{HTML}{228B22}
\definecolor{urlblue}{HTML}{000000}

\newcommand{\ie}{i.e.\xspace}
\newcommand{\eg}{e.g.\xspace}

\mathchardef\mhyphen="2D

\newcommand{\roughly}{\ensuremath{ {\sim}\,} }

\newlength{\dhatheight}

\newcommand{\code}[1]{\texttt{#1}\xspace}

\newcommand{\unit}[1]{\ensuremath{\mathrm{\,#1}}\xspace}

\newcommand{\Gyr}{\unit{Gyr}}

\newcommand{\pc}{\unit{pc}}
\newcommand{\kpc}{\unit{kpc}}

\newcommand{\Msun}{\unit{M_\odot}}

\newcommand{\Dsun}{\unit{D_\odot}}

\newcommand{\magn}{\unit{mag}}
\newcommand{\mmag}{\unit{mmag}}

\newcommand{\Mv}{\ensuremath{M_{V}}\xspace}

\newcommand{\secref}[1]{Section~\ref{sec:#1}}

\newcommand{\tabref}[1]{Table~\ref{tab:#1}}

\newcommand{\figref}[1]{Figure~\ref{fig:#1}}

\newcommand{\bandvar}[2][]{%
  \ifthenelse{\isempty{#1}}{\var{#2}}{\var{#2\_#1}}%
}

\newcommand{\modulus}{\ensuremath{m - M}\xspace}

\newcommand{\ra}{{\ensuremath{\alpha_{2000}}}\xspace}
\newcommand{\dec}{{\ensuremath{\delta_{2000}}}\xspace}
\newcommand{\age}{{\ensuremath{\tau}}\xspace}
\newcommand{\metal}{{\ensuremath{Z}}\xspace}

\newcommand{\ellip}{\ensuremath{\epsilon}\xspace}
\newcommand{\PA}{\ensuremath{\mathrm{P.A.}}\xspace}
\newcommand{\TS}{\ensuremath{\mathrm{TS}}\xspace}

\newcommand{\SExtractor}{\code{SExtractor}}

\newcommand{\HEALPix}{\code{HEALPix}}
\newcommand{\healpix}{\HEALPix}

\newcommand{\emcee}{\code{emcee}}
\newcommand{\ugali}{\code{ugali}}
\newcommand{\var}[1]{\ensuremath{\texttt{\MakeUppercase{#1}}}\xspace}

\providecommand\physrep{\ref@jnl{Phys.~Rep.}}%
\providecommand\apjs{\ref@jnl{ApJS}}%
\providecommand{\jcap}{\ref@jnl{JCAP}}%

\begin{document}

\reportnum{\footnotesize FERMILAB-PUB-18-653-AE}

\title{A faint halo star cluster discovered in the Blanco Imaging of the Southern Sky Survey}


\author{S. Mau}
\affiliation{Kavli Institute for Cosmological Physics, University of Chicago, Chicago, IL 60637, USA}
\affiliation{Department of Astronomy and Astrophysics, The University of Chicago, Chicago, IL 60637, USA}
\author{A.~Drlica-Wagner}
\affiliation{Fermi National Accelerator Laboratory, P. O. Box 500, Batavia, IL 60510, USA}
\affiliation{Kavli Institute for Cosmological Physics, University of Chicago, Chicago, IL 60637, USA}
\affiliation{Department of Astronomy and Astrophysics, The University of Chicago, Chicago IL 60637, USA}
\author{K.~Bechtol}
\affiliation{Department of Physics, University of Wisconsin, Madison, WI 53706, USA}
\author{A.~B.~Pace}
\affiliation{George P. and Cynthia Woods Mitchell Institute for Fundamental Physics and Astronomy, and Department of Physics and Astronomy, Texas A\&M University, College Station, TX 77843, USA}
\author{T.~Li}
\affiliation{Fermi National Accelerator Laboratory, P. O. Box 500, Batavia, IL 60510, USA}
\author{M.~Soares-Santos}
\affiliation{Department of Physics, Brandeis University, 415 South Street, Waltham, MA 02453, USA}
\author{N.~Kuropatkin}
\affiliation{Fermi National Accelerator Laboratory, P. O. Box 500, Batavia, IL 60510, USA}
\author{S.~Allam}
\affiliation{Fermi National Accelerator Laboratory, P. O. Box 500, Batavia, IL 60510, USA}
\author{D.~Tucker}
\affiliation{Fermi National Accelerator Laboratory, P. O. Box 500, Batavia, IL 60510, USA}
\author{L.~Santana-Silva}
\affiliation{Valongo Observatory, Federal University of Rio de Janeiro, Ladeira Pedro Antonio, 43, Rio de Janeiro, CEP 20080-090, Brazil}
\affiliation{NASA Goddard Space Flight Center, Greenbelt, MD 20771, USA}
\author{B.~Yanny}
\affiliation{Fermi National Accelerator Laboratory, P. O. Box 500, Batavia, IL 60510, USA}
\author{P.~Jethwa}
\affiliation{European Southern Observatory, Karl-Schwarzchild-Str. 2, D-85748, Garching, Germany}
\author{A.~Palmese}
\affiliation{Fermi National Accelerator Laboratory, P. O. Box 500, Batavia, IL 60510, USA}
\author{K.~Vivas}
\affiliation{Cerro Tololo Inter-American Observatory, National Optical Astronomy Observatory, Casilla 603, La Serena, Chile}
\author{C.~Burgad}
\affiliation{Department of Physics \& Astronomy, Ohio University, Clippinger Lab, Athens, OH 45701, USA}
\author{H.-Y.~Chen}
\affiliation{Black Hole Initiative, Harvard University, Cambridge, MA 02138, USA}

\email{sidneymau@uchicago.edu; kadrlica@fnal.gov}

\collaboration{(BLISS Collaboration)}

\email{}

\begin{abstract}
We present the discovery of a faint, resolved stellar system, \juicy, found in Dark Energy Camera data from the first observing run of the Blanco Imaging of the Southern Sky (\bliss) Survey.
\juicy is located at $(\ra, \dec) = (177\fdg511, -41\fdg772)$ with a heliocentric distance of $\Dsun = 23.7^{+1.9}_{-1.0} \kpc$.
It is a faint, $\Mv = 0.0^{+1.7}_{-0.7} \magn$, and compact, $r_h = 4.1^{+1}_{-1} \pc$, system consistent with previously discovered faint halo star clusters. 
Using data from the second data release of the \Gaia satellite, we measure a proper motion of $(\mu_{\alpha} \cos \delta, \mu_{\delta}) = (-2.37 \pm 0.06, 0.16 \pm 0.04)$ mas/yr.
Combining the available positional and velocity information with simulations of the accreted satellite population of the Large Magellanic Cloud (LMC), we find that it is unlikely that \juicy originated with the LMC.
\end{abstract}


\section{Introduction}
\label{sec:intro}

Resolved stellar systems offer a unique way to probe hierarchical structure formation at the outer reaches of the Milky Way halo \citep[e.g.,][]{Bullock:2005}.
The stellar populations of Milky Way satellites provide a statistical means to estimate ages, distances, and metallicities from photometry alone, while also providing a coherent tracer of the Milky Way's gravitational potential at large distances.
Over the past several years, the search for dark-matter-dominated ultra-faint satellite galaxies has received much attention \citep[\eg,][]{McConnachie:2012,Bechtol:2015wya,Koposov:2015cua,Drlica-Wagner:2015ufc}.
Ultra-faint galaxies are characterized by their large physical sizes, relative to their low luminosities, and their large stellar velocity dispersions, relative to their small observed stellar masses. 
The discovery of ultra-faint galaxies has come hand-in-hand with the discovery of a class of ultra-faint halo star clusters \citep[e.g.,][]{Fadely:2011,Munoz:2012,Balbinot:2013,Belokurov:2014,Laevens:2014,Kim:2015a,Laevens:2015b,Kim:2016,Luque:2016,Luque:2017,Koposov:2017,Luque:2018,Torrealba:2019}.
These clusters have physical sizes consistent with the population of globular clusters (a few parsecs), but can have luminosities hundreds of times fainter ($M_V \gtrsim -2.5$).
It has been proposed that halo clusters were accreted onto the Milky Way through the disruption of infalling satellite galaxies \citep[\eg,][]{Searle:1978, Gnedin:1997, Koposov:2007, Forbes:2010, Leaman:2013, Massari:2017}.
This scenario has received considerable observational support from the age, metallicity, and spatial distribution of these clusters \citep[\eg,][]{Zinn:1993a, DaCosta:1995, Marin-Franch:2009, Mackey:2004, Mackey:2010, Dotter:2010, Keller:2011}.
The accretion scenario is further supported by the close resemblance between Milky Way halo clusters and the clusters of dwarf galaxies associated with the Milky Way \citep[\eg,][]{Smith:1998, Johnson:1999, DaCosta:2003, Wetzel:2015, Yozin:2015, Bianchini:2017}.

Over the last several years, there has been renewed interest in the possibility that a significant fraction of the Milky Way's ultra-faint satellites may have been accreted with the Large Magellanic Cloud \citep[LMC;][]{Deason:2015, Jethwa:2016}.
Proper motion measurements with \Gaia have significantly increased the likelihood that several ultra-faint galaxies are associated with the LMC \citep[\eg,][]{Kallivayalil:2018}.
Understanding the relationship between new ultra-faint systems and the LMC will help quantify the influence of the LMC on the Milky Way satellite population.

In this context, we report the discovery of a new faint outer-halo star cluster, \juicy, using multi-band imaging data from the Dark Energy Camera (DECam) as part of the Blanco Imaging of the Southern Sky (\bliss) survey.
In \secref{data} we describe the \bliss data and processing along with our use of \Gaia data, while in \secref{search} we describe our search for resolved stellar systems.
The systemic and structural properties of \juicy are reported in \secref{results}. 
In \secref{discussion} we compare the measured properties of \juicy to other known faint halo star clusters, and we conclude by discussing possible origins of this faint stellar system.

\begin{figure*}[t]
\center
\includegraphics[width=\textwidth]{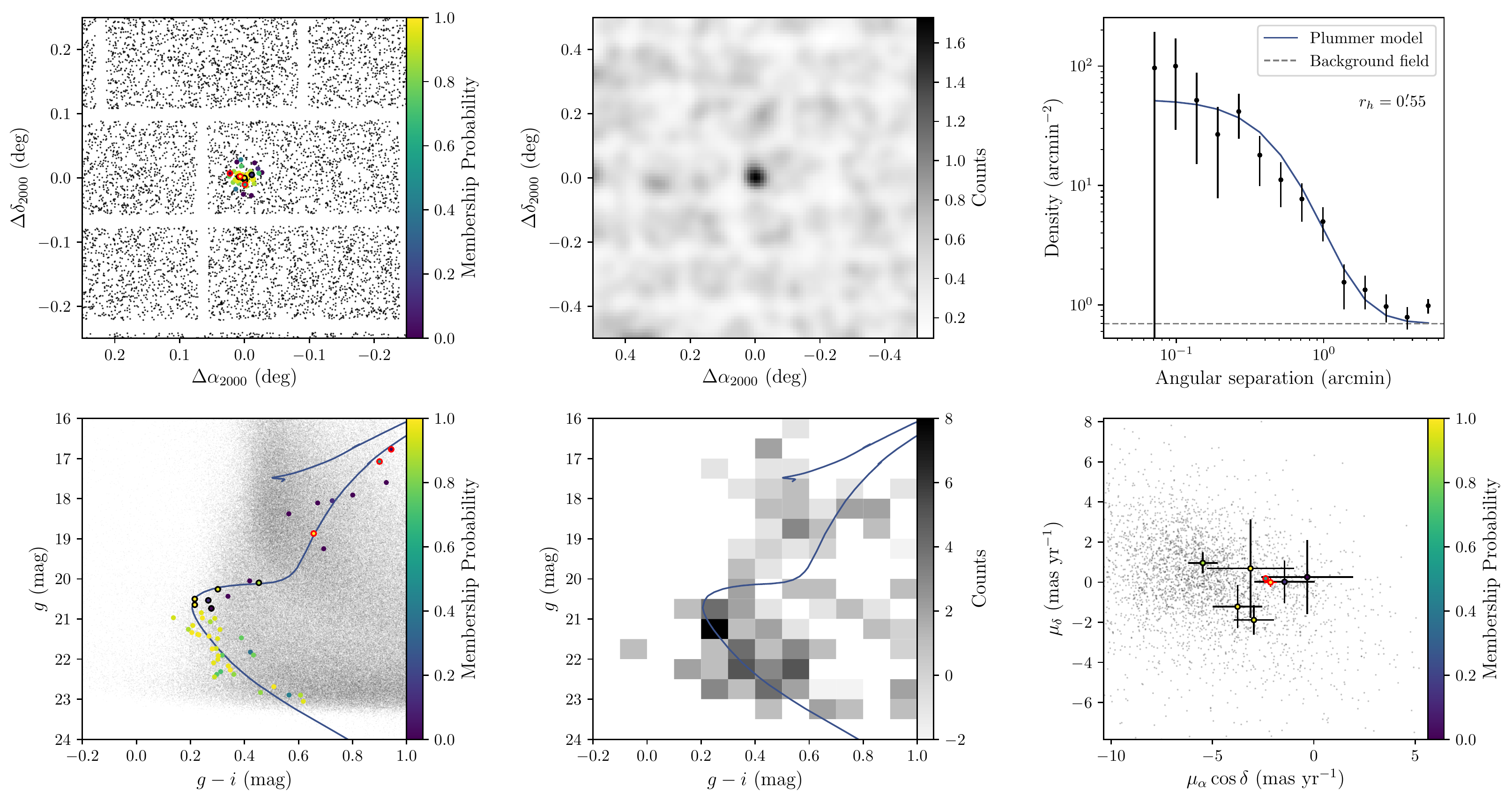}
\caption{
    Stellar density, color-magnitude diagrams (CMDs), radial distribution, and proper motions for \juicy. 
    Stars with $g < 24.5\magn$ are colored by their best-fit \ugali membership probability, $p$ (\secref{results}); objects with $p < 5\%$ are shown in gray.
    Top left: spatial distribution of stars passing the isochrone filter.
    Top center: binned spatial distribution of stars passing the isochrone filter, smoothed by a Gaussian kernel with $\sigma = 2\farcm2$.
    Top right: radial distribution of isochrone-filtered stars with respect to the centroid of \juicy.
    The blue curve corresponds to the best-fit Plummer model, assuming spherical symmetry, with $r_h = 0\farcm55$.
    The dashed gray line represents the background field density.
    Bottom left: the CMD of stars within one half-light radius of the centroid; the blue curve corresponds to the best-fit isochrone.
    The distribution of background stars in the surrounding region is shown in gray.
    Bottom center: background-subtracted Hess diagram for stars within $3 r_h$ of the centroid of \juicy.
    The background is estimated from a concentric annulus with $5 r_h < r < 5.8 r_h$.
    The best-fit isochrone is overplotted in blue.
    Bottom right: \Gaia proper motions for stars cross-matched with \bliss (\secref{data}).
    Color represents the \ugali membership probability derived from the \bliss data.
    The proper motions of field stars are shown in gray.
}
\label{fig:figure_1}
\end{figure*}

\section{Data}
\label{sec:data}

The \bliss survey is a multiband, multipurpose program that uses DECam \citep{Flaugher:2015} to image the southern sky in the $g$, $r$, $i$, and $z$ bands.\footnote{We note that the $z$-band data is not used in this analysis.}
\bliss processes all public DECam exposures with exposure time $>30$ s in addition to performing devoted observations to fill gaps in the existing DECam sky coverage.
The first \bliss observing campaign consisted of 11 nights in 2017A (Prop-ID: 2017A-0386). 
DECam exposures were reduced and processed by the Dark Energy Survey (DES) Data Management system using the same pipeline that was applied to the DES public data release \citep[DES DR1;][]{DES:2018, Morganson:2018}.
Astronomical source detection and photometry were performed on a per exposure basis using the \code{PSFex} and \code{SExtractor} routines \citep{Bertin:1996, Bertin:2011}. 
As part of this step, astrometric calibration was performed with \code{SCAMP} (Software for Calibrating AstroMetry and Photometry) \citep{Bertin:2006} by matching objects to the 2MASS catalog \citep{Skrutskie:2006}. 
The \code{SExtractor} source detection threshold was set to detect sources with  ${\rm S/N} \gtrsim 5$.
The photometric fluxes and magnitudes used here refer to the \SExtractor PSF model fit.

Photometric calibration was performed by matching stars to the APASS \citep{Henden:2014} and 2MASS \citep{Skrutskie:2006} catalogs for each CCD (charge-coupled device) following the procedure described in \citet{Drlica-Wagner:2016}.
Briefly, APASS-measured magnitudes were transformed to the DES system before calibration, following the equations described in Appendix A4 of \citet{Drlica-Wagner:2018}:
\small\begin{align*}
g_{\rm DES} &= g_{\rm APASS} - 0.0642(g_{\rm APASS}-r_{\rm APASS}) - 0.0239 \\
r_{\rm DES} &= r_{\rm APASS} - 0.1264(r_{\rm APASS}-i_{\rm APASS}) - 0.0098 \\
i_{\rm DES} &= r_{\rm APASS} - 0.4145(r_{\rm APASS}-J_{\rm 2MASS} - 0.81) - 0.0391,
\end{align*}\normalsize
which have statistical root-mean-square errors per star of $\sigma_g = 0.04\magn$, $\sigma_i = 0.04\magn$, and $\sigma_r = 0.05\magn$.
For a small number of CCDs where too few stars were matched, or the resulting zeropoint was a strong outlier with respect to the other CCDs in that exposure, zeropoints were instead derived from a simultaneous fit to all CCDs in the exposure.
This calibration procedure is similar to that described by \citet{Koposov:2015cua}, and the relative photometric uncertainty in the derived zeropoints is estimated to be $\sim 3\%$.
Extinction from interstellar dust was calculated for each object from a bilinear interpolation (in Galactic coordinates) of the extinction maps of \citet{Schlegel:1998}.
To calculate reddening, we assumed $R_V = 3.1$ and used a set of $R_b = A_b/E(B-V)$ coefficients derived by DES for the $g$, $r$, and $i$ bands: $R_g = 3.185$, $R_r = 2.140$, and $R_i = 1.571$ \citep{DES:2018}.\footnote{An update to the DECam standard bandpasses changed these coefficients by $< 1$ \mmag for DES DR1 \citep{DES:2018}.}
All quoted magnitudes have been de-reddened using this procedure.

A multiband catalog of unique objects was assembled by performing a 1\arcsec\ match for all objects detected in each of the individual exposures following the procedure described in \citet{Drlica-Wagner:2015ufc}.
In the region surrounding \juicy, the median $10 \sigma$ limiting depth of the \bliss catalog is $g \sim 23.0 \magn$ and $i \sim 21.8 \magn$. 
We selected stellar objects based on the \var{spread\_model} quantity: $|\var{spread\_model\_g}| < 0.003 \allowbreak + \var{spreaderr\_model\_g}$.
This selection has been shown to yield a stellar completeness of $\roughly 90\%$ to a magnitude of $g \sim 23 \magn$ \citep{Drlica-Wagner:2015ufc}.

Our search for resolved stellar systems in the \bliss data uses $g$- and $r$-band photometry.  
However, \juicy was identified in a region where the \bliss $r$-band data did not provide full coverage at consistent depth, leading us to use $i$ instead of $r$ for parameter fitting (\ie, in \secref{results} and \figref{figure_1}).
The $i$ band is slightly shallower than the $r$ band, but we gain full coverage in the region surrounding \juicy, allowing us to better estimate the distribution of field stars.

For our proper motion analysis, we spatially cross-match stars in the \bliss catalog to objects in \Gaia DR2 \citep{Gaia:2018} using a matching radius of $0\farcs5$.
We remove Milky Way foreground stars using a parallax cut of $\varpi - 3 \sigma_{\varpi} > 0$ \citep{Pace:2018}.
Defining $u = (\code{astrometric\_chi2\_al} /  \allowbreak (\code{astrometrc\_n\_good\_obs\_al} - 5))^{1/2}$, we remove stars with bad astrometric fits of $u > 1.2 \times \max\{1, \exp(-0.2 (G - 19.5))\}$ \citep{Lindegren:2018, Pace:2018}.
We note that the cross-match with \Gaia and the subsequent cuts based on \Gaia-measured quantities are only applied when deriving the proper motion of \juicy in \secref{motion}.

\begin{figure*}[t]
\center
\includegraphics[width=\textwidth]{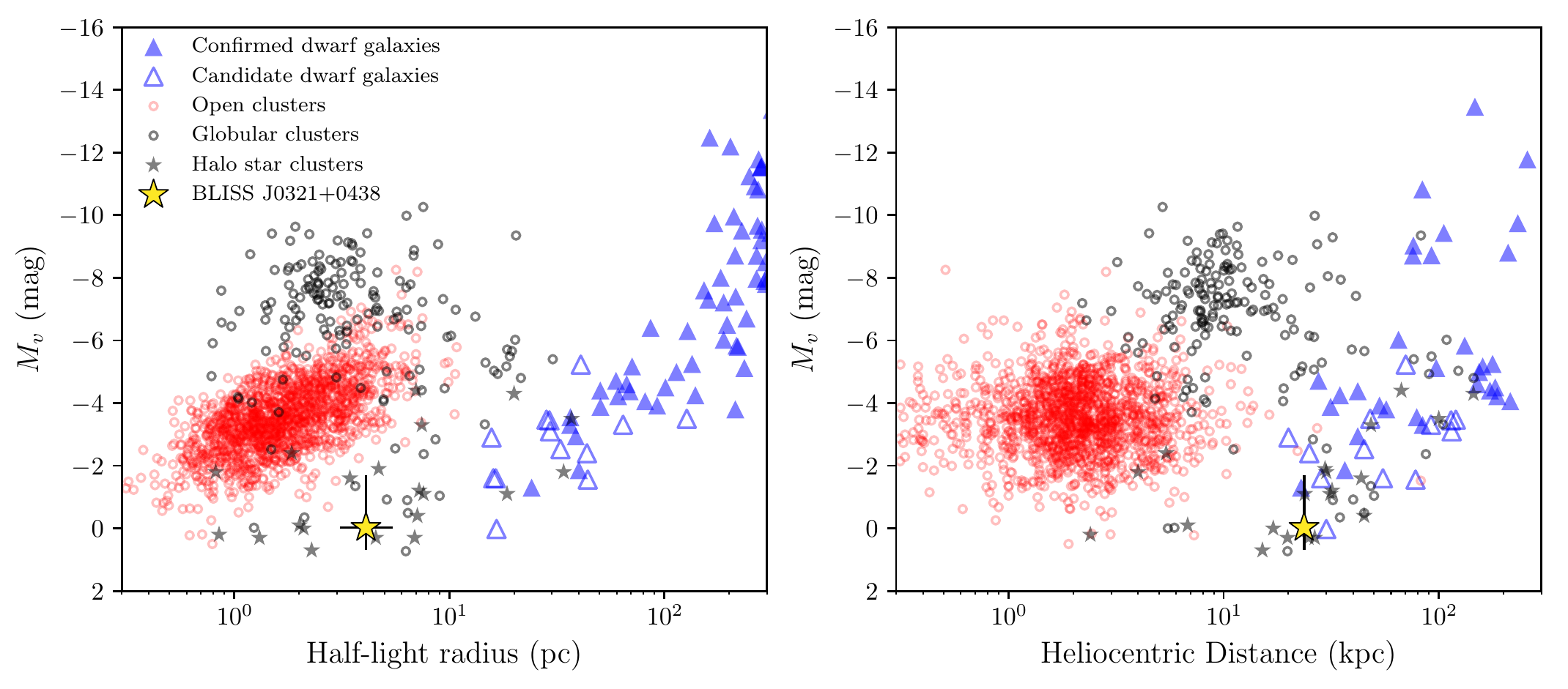}
\caption{
    Physical size vs.\ absolute magnitude (left), and heliocentric distance vs.\ absolute magnitude (right) for resolved stellar systems in and around the Milky Way.
    Milky Way globular clusters are shown as black circles \citep{Harris:1996}, Milky Way open clusters are shown as red circles \citep{Kharchenko:2013}, and recently discovered faint halo star clusters are shown as black stars. Local Group dwarf galaxies are shown as solid blue triangles, while candidate Milky Way satellite galaxies without firm classification are shown as open blue triangles \citep[][and references therein]{Torrealba:2018b}.
}
\label{fig:figure2}
\end{figure*}

\section{Satellite Search}
\label{sec:search}

We perform a matched-filter search for spatial overdensities of old, metal-poor stars in the \bliss data by applying an algorithm inspired by the binned stellar density map procedure described in Section 3.1 of \citet{Bechtol:2015wya}.
We partition the sky into \healpix pixels of nside=32 ($\roughly 3.4 \deg^2$) \citep{Gorski:2005}.
Iterating through all \healpix pixels, we select \bliss stellar catalogs in each \healpix pixel and its eight nearest-neighbor \healpix pixels.
We apply quality cuts requiring that objects be detected in both $g$ and $r$ bands and are brighter than $g = 24.5 \magn$ ($\code{WAVG\_MAG\_PSF\_G} < 24.5$).
We filter the stellar catalogs according to a template \code{PARSEC} isochrone \citep{Bressan:2012} with age $\age=12 \Gyr$ and metallicity $\metal=0.0001$.
We scan this isochrone over a range of distance moduli from $16 \leq \modulus \leq 24.5$ in steps of $0.5\magn$.
At each distance modulus step, we select stars consistent with the template isochrone by requiring the color difference between each star and the template isochrone to be $\Delta (g-r) < \sqrt{0.1^2 + \sigma_g^2 + \sigma_r^2}$, where $\sigma_g$ and $\sigma_r$ are the statistical uncertainties on the $g$- and $r$-band magnitudes, respectively.
For each of these isochrone filters, we compute the characteristic density of the data in the region of interest and convolve the stellar field with a Gaussian kernel of $2 \arcmin$ to find peaks relative to the mean field density of the \healpix pixel.
For each peak, we compute the characteristic local density in an annulus around the peak, fitting an aperture that yields the highest Poisson significance relative to the background.
To fit these apertures, we vary the radius of an annulus about each peak, and compute the Poisson significance of objects inside, with respect to the expected model number from the local characteristic density.
We remove redundant peaks (\ie, spatially coincident peaks at different distance moduli) and compile a candidate list.

We make sets of diagnostic plots for all candidates with Poisson significance ${>}\,5.5 \sigma$ and examine each by eye.
We compare this list of candidates against previously reported systems to identify new and unique objects. 
\juicy is detected with a Poisson significance of ${>}\,14 \sigma$ and is the first significant candidate discovered in this search .

\begin{figure*}[t]
\center
\includegraphics[width=\textwidth]{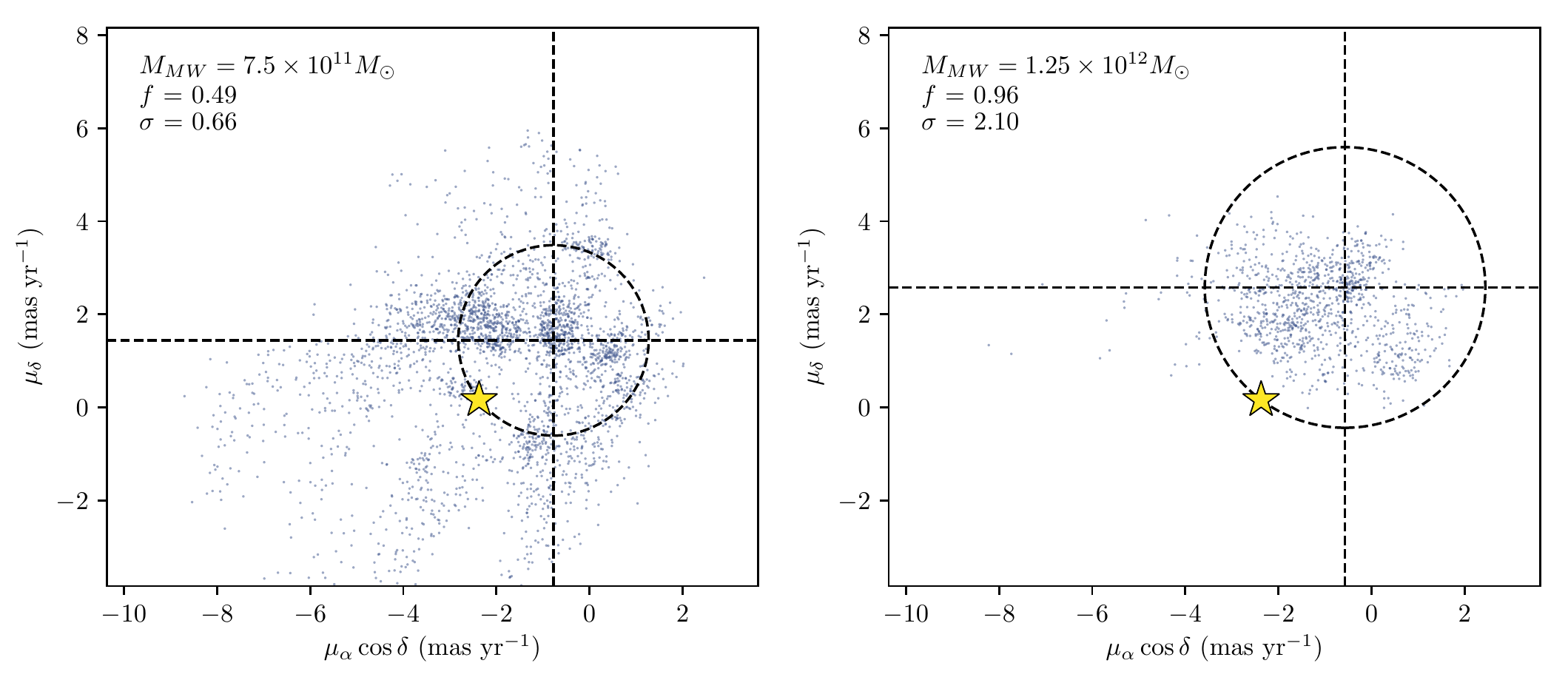}
\caption{
    Proper motion distribution of simulated LMC satellite particles within 10\kpc of \juicy from \citet{Jethwa:2016}. 
    The proper motion of \juicy is represented as a yellow star (uncertainty in the proper motion is dwarfed by the scale of the axes).
    The dashed lines indicate the mode of the simulated satellite distribution, while the dashed circle encloses points that are closer to the mode than is \juicy.
    The left panel assumes $M_{\rm MW} = 7.5 \times 10^{11} \Msun$, while the right panel assumes $M_{\rm MW} = 1.25 \times 10^{12} \Msun$. 
    In both cases the LMC mass is set to  $M_{\rm LMC} = 2 \times 10^{11} \Msun$.
}
\label{fig:figure3}
\end{figure*}

\section{Results}
\label{sec:results}

After identifying \juicy as a statistically significant overdensity, we proceed to perform fits to the morphological, isochrone, and proper motion parameters of this system.
We describe these fits in \secref{morph} and \secref{motion}, and we present the results in \figref{figure_1} and \tabref{properties}. 

\subsection{Morphological and Isochrone Parameters}
\label{sec:morph}

We fit the morphological and isochrone parameters of \juicy using the maximum likelihood formulation implemented in \ugali \citep{Bechtol:2015wya,Drlica-Wagner:2015ufc}. 
We model the spatial distribution of stars with a Plummer profile and the CMD with a synthetic isochrone from \citet{Bressan:2012}. 
We simultaneously fit the longitude, latitude, extension, ellipticity, and position angle of the Plummer profile, and the age, metallicity, and distance modulus of the isochrone.
We derive the posterior probability distribution using an affine invariant Markov Chain Monte Carlo (MCMC) ensemble sampler  \citep[i.e., \code{emcee}; ][]{Foreman-Mackey:2013}.
The resulting best-fit parameters are shown in \tabref{properties}. 
Three-dimensional physical separations between \juicy and the Galactic Center, LMC, and SMC are reported in \tabref{distances}.
\juicy is significantly detected in our likelihood analysis with a test statistic of ${\rm TS} = 206.7$, corresponding to a Gaussian significance of $14.4 \sigma$.
We find that \juicy is an old ($\age = 9.2\Gyr$), compact ($r_h = 0\farcm58$) stellar system with a higher average systemic metallicity ($\metal = 0.0006$) than most ultra-faint dwarf galaxies \citep[][2015 edition]{McConnachie:2012}.
In the left panels of \figref{figure_1}, we plot the spatial (top) and color-magnitude (bottom) distributions of stars in the region surrounding \juicy. 
Stars are colored according to their membership probability, as determined by the likelihood fit (stars with $p < 5\%$ are colored gray).
The center panels further support the reality of this system by showing the binned spatial distribution of objects passing the isochrone filter of $\Delta (g-i) < \sqrt{0.1^2 + \sigma_g^2 + \sigma_i^2}$, where $\sigma_g$ and $\sigma_i$ are the statistical uncertainties on the $g$- and $i$-band magnitudes (top), and the background-subtracted Hess diagram of objects within $3 r_h$ of the best-fit centroid (bottom).
The top right panel shows the radial distribution of objects passing the isochrone filter with respect to the centroid of \juicy; the distribution is consistent with what we expect from the top left and top center panels and is well-modeled by a Plummer profile.

The physical parameters of \juicy place it in a somewhat ambiguous region of parameter space. 
It is fainter than most classical globular clusters, more distant than most known open clusters, and more compact than ultra-faint satellite galaxies.
As such, \juicy appears to belong to a poorly defined group of old, metal-poor systems that resemble globular clusters, but are fainter than conventional globular clusters \citep[][2010 edition]{Harris:1996}.
Based on these properties, we tentatively classify \juicy as an ultra-faint halo star cluster rather than an ultra-faint dwarf galaxy.
Spectroscopy is necessary to confirm this classification through systemic velocity and metallicity dispersion measurements.

\subsection{Proper Motion}
\label{sec:motion}

We cross-match \bliss data with \Gaia to search for correlated proper motion in the stars identified as high-probability members of \juicy.
Because \Gaia has a limiting magnitude of $g \approx 21\magn$ \citep{Gaia:2018}, we are only sensitive to the proper motions of the brighter members of \juicy when cross-matching.
We estimate the proper motion of \juicy by applying the Gaussian mixture model analysis described in Section~2.2 of \citet{Pace:2018} to all cross-matched stars.
Briefly, the mixture model separates the likelihoods of the satellite and the Milky Way stars, decomposing each into a product of spatial and proper-motion likelihoods.
Stars that are closer to the centroid of \juicy are weighted more heavily by assuming the best-fit projected Plummer profile.
This analysis yields a proper motion for \juicy of $(\mu_{\alpha} \cos \delta, \mu_{\delta}) = (-2.37^{+0.06}_{-0.06}, 0.16^{+0.04}_{-0.04})$ mas yr${}^{-1}$.

The brightest red giant branch (RGB) stars are precisely measured by \Gaia, and we find that their proper motions are tightly clustered (red points in the lower right panel of \figref{figure_1}).
To cross-check the results of the more sophisticated mixture model, we use a simple MCMC to determine the average proper motion, including correlated uncertainties, of just the three candidate RGB members of \iauname\xspace (\bliss 1; circled red in \figref{figure_1}).
Doing this, we find $(\mu_{\alpha} \cos \delta, \mu_{\delta}) = (-2.36^{+0.06}_{-0.06}, 0.17^{+0.04}_{-0.05})$ mas yr${}^{-1}$.
This simple cross-check is in good agreement with the more comprehensive mixture model analysis, and we quote the mixture model result as our primary proper motion measurement.
The correlated proper motions of the brightest likely-member stars lends confidence to the claim that \juicy is a real satellite system.

\section{Discussion}
\label{sec:discussion}

\juicy joins the rapidly growing list of ultra-faint halo star clusters discovered in recent years with the advent of DES, Pan-STARRS, \Gaia, and other wide-area surveys. 
In \figref{figure2} we compare \juicy to known satellite galaxies, globular clusters, and open clusters in the space of absolute magnitude versus physical half-light radius, and absolute magnitude versus heliocentric distance.\footnote{
We follow the analysis of \citet{Torrealba:2019} to derive values for open clusters in \citet{Kharchenko:2013}.
Half-light radii are estimated as the radius at which the integrated King profile (Eq.~2 from \citealt{Piskunov:2007}) is half the value integrated to the tidal radius.
Luminosities are estimated by fitting an isochrone to member stars from the $1 \sigma$ membership probability group described in \citet{Piskunov:2007}.
}
\juicy is significantly fainter than classical globular clusters \citep[][2010 edition]{Harris:1996}, more distant than most open clusters \citep{Kharchenko:2013}, and more compact than recently discovered ultra-faint dwarf galaxies \citep[\eg,][ and references therein]{Torrealba:2018b}.
The size and luminosity of \juicy make it similar to recently discovered faint halo clusters, such as Koposov~1 and 2 \citep{Koposov:2007}, Kim~1 \citep{Kim:2015a}, Segue~3 \citep{Fadely:2011}, Mu{\~n}oz~1 \citep{Munoz:2012}, Kim~3 \citep{Kim:2016}, and the nine objects discovered by \citet{Torrealba:2019}. 
These objects populate the ``valley of ambiguity'' \citep{Gilmore:2007} between compact star clusters and extended dwarf galaxies.

\juicy resides in a region of the sky where the gravitational influence of the LMC is non-negligible. 
While \juicy is currently located 40.5\kpc ($53\fdg5$ in projection) from the LMC, it is natural to consider whether \juicy may have originated with the LMC before being tidally captured by the Milky Way during accretion of the LMC. 
We test this hypothesis by comparing with numerical simulations from \citet{Jethwa:2016}, which model the evolution of the LMC satellite population during infall.
These simulations sample plausible ranges of LMC orbital histories and initial total masses for the Milky Way ($M_{\rm MW}$) and LMC ($M_{\rm LMC}$).
We estimate the statistical consistency between the proper motion of \juicy and the simulated distribution of satellites within 10\kpc of its location (\figref{figure3}). 
We calculate a $p$-value from the fraction of simulated satellites that are farther from the mode of the simulated proper motion distribution than \juicy and convert this $p$-value to a Gaussian significance.
We find that the distribution of simulated satellites within 10\kpc of \juicy has a predicted heliocentric proper motion of $(\mu_{\alpha} \cos \delta, \mu_{\delta}) = (-1.56, 2.61)$ mas yr${}^{-1}$ for $M_{\rm MW} = 7.5 \times 10^{11} \Msun$ (consistent with \citealt{Eilers:2018}) and  $M_{\rm LMC} = 2 \times 10^{11} \Msun$ (consistent with \citealt{Penarrubia:2016}).
The proper motion of \juicy has $p=0.35$ and is consistent with the simulated proper motion of the LMC debris at the level of $0.66 \sigma$.
For $M_{\rm MW} = 1.25 \times 10^{12} \Msun$ (consistent with \citealt{Watkins:2018}), the simulated distribution of satellites has a predicted heliocentric proper motion of $(\mu_{\alpha} \cos \delta, \mu_{\delta}) = (-1.52, 3.51)$ mas yr${}^{-1}$.
For this distribution, \juicy has $p=0.03$ and is consistent with the proper motion of the simulated satellites at the level of $2.10 \sigma$.

\begin{figure}[h]
\center
\includegraphics[width=\columnwidth]{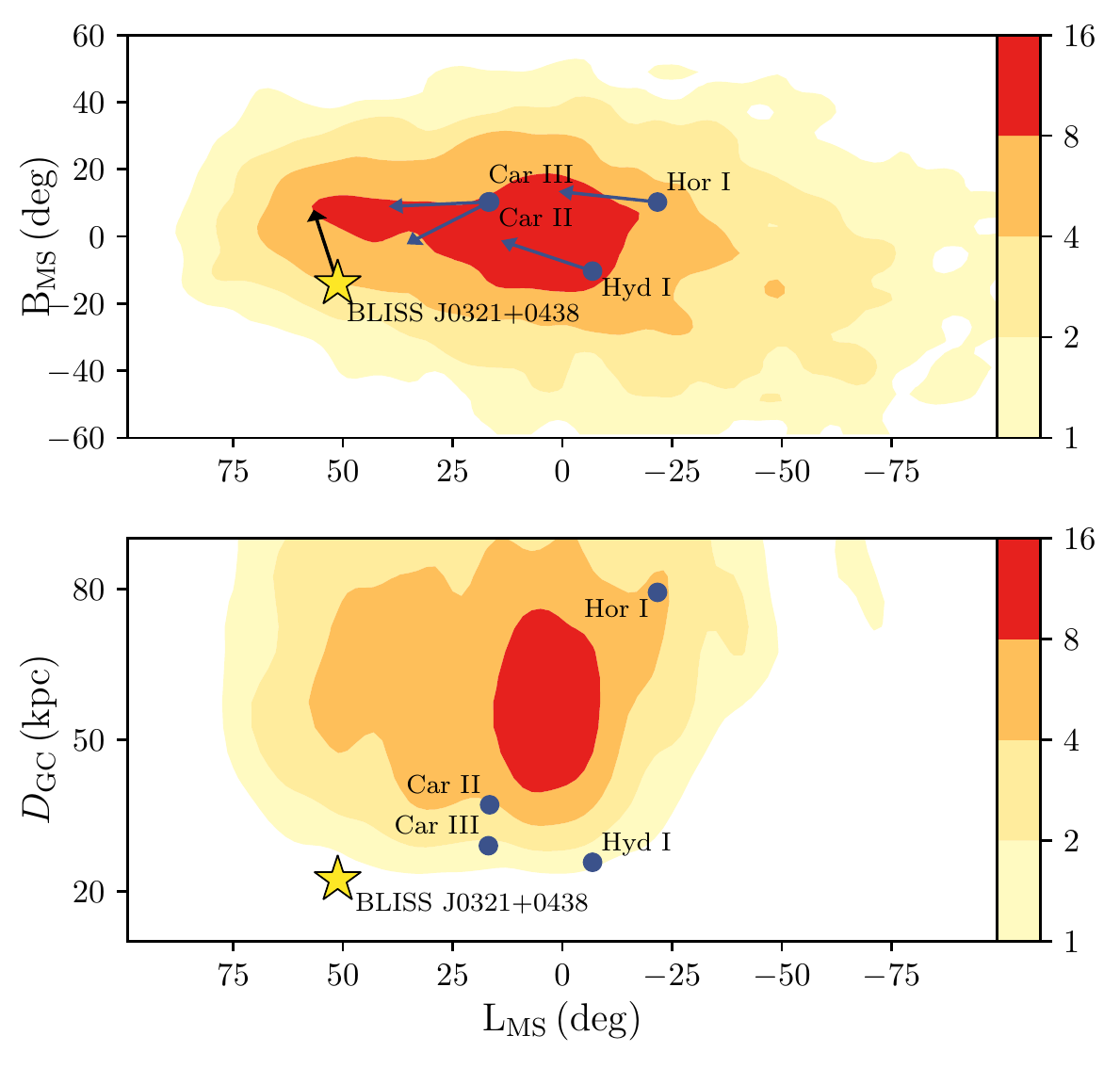}
\caption{
    Normalized relative density of simulated LMC satellites from \citet{Jethwa:2016}.
    \juicy is shown as a yellow star, and the four satellites of likely LMC origin \citep[Hor I, Car II, Car III, and Hyd I;][]{Kallivayalil:2018} are shown as blue dots.
    The color scale corresponds to the normalized relative density of LMC satellites in arbitrary units.
    Upper: spatial position of \juicy and the simulated LMC tidal debris.
    Arrows indicate the solar-reflex-corrected proper motions of \juicy and the four satellites with likely LMC origins (no physical meaning is attributed to the magnitudes of these arrows).
    Note that Car~II and Car~III are nearly coincident but have different velocity vectors.
    Lower: Galactocentric distance vs.\ Magellanic Stream longitude.
    \juicy is closer to the Galactic center than most of the simulated LMC tidal debris.
    In comparison, the four satellites with likely LMC origins are more consistent with the simulated distribution.
}
\label{fig:figure4}
\end{figure}

As a visual cross-check, we convert the solar-reflex-corrected proper motions of \juicy to the Magellanic Stream coordinate system \citep{Nidever:2008} and plot the relative velocity vector, along with similar vectors for several satellites with probable LMC origins \citep{Kallivayalil:2018}, in the upper panel of \figref{figure4}.
The lower panel of \figref{figure4} shows the Galactocentric distance versus Magellanic Stream longitude.
The proper motion of \juicy is not preferentially aligned with the on-sky distribution of simulated LMC debris, nor is it aligned with the velocities of satellites with probable LMC origins.
It is possible that \juicy was instead accreted with another satellite galaxy progenitor that has already been tidally disrupted by the Milky Way.
It is also possible that \juicy has a common origin with other Milky Way open or globular clusters, and that it is merely an outlier in distance and luminosity.

The rapidly growing catalog of faint outer-halo star clusters \citep[e.g.,][]{Fadely:2011,Munoz:2012,Balbinot:2013,Belokurov:2014,Laevens:2014,Laevens:2015b,Kim:2015a,Kim:2016,Luque:2016,Luque:2017,Luque:2018,Koposov:2017,Torrealba:2019}, emphasizes the incompleteness of current surveys.
As deep, homogeneous sky coverage improves (\ie, with the Large Synoptic Survey Telescope (LSST)), it is likely that a much larger number of similar faint, small systems will be discovered.
Characterizing these faint objects will become increasingly important for understanding the formation and origins of resolved stellar systems in the Milky Way halo.


\begin{deluxetable}{c c c}
\tablecolumns{3}
\tablewidth{0pt}
\tabletypesize{\footnotesize}
\tablecaption{\label{tab:properties}
Observed and derived properties of \juicy.
}
\tablehead{
Parameter & Value & Units
}
\startdata
\ra & $177.510^{+0.003}_{-0.003}$ & degree \\
\dec & $-41.772^{+0.002}_{-0.002}$ & degree \\
$a_h$ & $0.60^{+0.19}_{-0.14}$ & arcmin \\
$r_h$ & $0.58^{+0.19}_{-0.14}$ & arcmin \\
$r_{1/2}$ & $4.1^{+1}_{-1}$ & pc \\
\ellip & $0.06$ & \ldots \\
\PA & $101^{+74}_{-56}$ & degree \\
\modulus & $16.87^{+0.17}_{-0.09} \pm 0.1$\tablenotemark{a} & \ldots \\
$D_{\odot}$ & $23.7^{+1.9}_{-1.0}$ & kpc \\
\age & $9.2^{+1.6}_{-0.8}$ & \Gyr \\
$\sum p_i$ & $58^{+16}_{-14}$ & \ldots \\
\TS & $206.713$ & \ldots \\[-0.5em]
\multicolumn{3}{c}{ \hrulefill } \\
$\mu_{\alpha} \cos \delta$ & $-2.37^{+0.06}_{-0.06}$ & mas yr${}^{-1}$ \\
$\mu_{\delta}$ & $0.16^{+0.04}_{-0.04}$ & mas yr${}^{-1}$ \\[-0.5em]
\multicolumn{3}{c}{ \hrulefill } \\
$M_V$ & $0.0^{+1.7}_{-0.7}$\tablenotemark{b} & mag \\
$M_{\star}$ & $143^{+34}_{-39}$ & $\Msun$ \\
$\mu$ & $26.1$ & mag arcsec${}^{-2}$ \\
${\rm [Fe/H]}$ & -1.4 & dex \\
$E(B-V)$ & 0.12 & \ldots \\[+0.5em]
\enddata
\tablecomments{Uncertainties were derived from the highest density interval containing the peak and 90\% of the marginalized posterior distribution.}
\tablenotetext{a}{We assume a systematic uncertainty of $\pm0.1$ associated with isochrone modeling.}
\tablenotetext{b}{The uncertainty in $M_V$ was calculated following \citet{Martin:2008} and does not include uncertainty in the distance.}
\vspace{-3em}
\end{deluxetable}

\begin{deluxetable}{c c c c c}
\hsize=\columnwidth
\tablecolumns{3}
\tablewidth{0pt}
\tabletypesize{\footnotesize}
\tablecaption{\label{tab:distances}
Derived physical separation of \juicy.
}
\tablehead{
Parameter & & Value & & Units
}
\startdata
$D_{\rm GC}$  & ~~~~~~~~ & 22.3 & ~~~~~~~~ & kpc \\
$D_{\rm LMC}$ & ~~~~~~~~ & 40.5 & ~~~~~~~~ & kpc \\
$D_{\rm SMC}$ & ~~~~~~~~ & 56.1 & ~~~~~~~~ & kpc \\[+0.5em]
\enddata
\tablecomments{Three-dimensional physical separation between \juicy and the Galactic Center, LMC, and SMC.}
\vspace{-3em}
\end{deluxetable}

\section{Acknowledgments}

We thank Gabriel Torrealba for sharing the table of open star clusters with estimated parameters.
S.M. is supported by the University of Chicago Provost's Scholar Award.
This project is partially supported by the NASA Fermi Guest Investigator Program Cycle 9 No.\ 91201. 


This project used data obtained with the Dark Energy Camera (DECam), 
which was constructed by the Dark Energy Survey (DES) collaboration.
Funding for the DES Projects has been provided by 
the DOE and NSF (USA),   
MISE (Spain),   
STFC (UK), 
HEFCE (UK), 
NCSA (UIUC), 
KICP (U. Chicago), 
CCAPP (Ohio State), 
MIFPA (Texas A\&M University),  
CNPQ, 
FAPERJ, 
FINEP (Brazil), 
MINECO (Spain), 
DFG (Germany), 
and the collaborating institutions in the Dark Energy Survey, which are
Argonne Lab, 
UC Santa Cruz, 
University of Cambridge, 
CIEMAT-Madrid, 
University of Chicago, 
University College London, 
DES-Brazil Consortium, 
University of Edinburgh, 
ETH Z{\"u}rich, 
Fermilab, 
University of Illinois, 
ICE (IEEC-CSIC), 
IFAE Barcelona, 
Lawrence Berkeley Lab, 
LMU M{\"u}nchen, and the associated Excellence Cluster Universe, 
University of Michigan, 
NOAO, 
University of Nottingham, 
Ohio State University, 
OzDES Membership Consortium
University of Pennsylvania, 
University of Portsmouth, 
SLAC National Lab, 
Stanford University, 
University of Sussex, 
and Texas A\&M University.

This work has made use of data from the European Space Agency (ESA) mission {\it Gaia} (\url{https://www.cosmos.esa.int/gaia}), processed by the {\it Gaia} Data Processing and Analysis Consortium (DPAC, \url{https://www.cosmos.esa.int/web/gaia/dpac/consortium}).
Funding for the DPAC has been provided by national institutions, in particular the institutions participating in the {\it Gaia} Multilateral Agreement.

Based on observations at Cerro Tololo Inter-American Observatory, National Optical Astronomy Observatory (2017A-0260; PI: Soares-Santos), which is operated by the Association of Universities for Research in Astronomy (AURA) under a cooperative agreement with the National Science Foundation.

This manuscript has been authored by Fermi Research Alliance, LLC under contract No.\ DE-AC02-07CH11359 with the U.S. Department of Energy, Office of Science, Office of High Energy Physics. The United States Government retains and the publisher, by accepting the article for publication, acknowledges that the United States Government retains a non-exclusive, paid-up, irrevocable, world-wide license to publish or reproduce the published form of this manuscript, or allow others to do so, for United States Government purposes.

\facility{Blanco, Gaia.}
\software{\code{Matplotlib} \citep{Hunter:2007},  \code{numpy} \citep{numpy:2011}, \code{scipy} \citep{scipy:2001}, \code{astropy} \citep{Astropy:2013}, \healpix \citep{Gorski:2005},\footnote{\url{http://healpix.sourceforge.net}} \code{healpy},\footnote{\url{https://github.com/healpy/healpy}} \code{fitsio},\footnote{\url{https://github.com/esheldon/fitsio}} \emcee \citep{Foreman-Mackey:2013}, \ugali \citep{Bechtol:2015wya}.\footnote{\url{https://github.com/DarkEnergySurvey/ugali}}}

\bibliography{main}

\begin{thebibliography}{}
\expandafter\ifx\csname natexlab\endcsname\relax\def\natexlab#1{#1}\fi
\providecommand{\url}[1]{\href{#1}{#1}}

\bibitem[{{Astropy Collaboration} {et~al.}(2013){Astropy Collaboration},
  {Robitaille}, {Tollerud}, {Greenfield}, {Droettboom}, {Bray}, {Aldcroft},
  {Davis}, {Ginsburg}, {Price-Whelan}, {Kerzendorf}, {Conley}, {Crighton},
  {Barbary}, {Muna}, {Ferguson}, {Grollier}, {Parikh}, {Nair}, {Unther},
  {Deil}, {Woillez}, {Conseil}, {Kramer}, {Turner}, {Singer}, {Fox}, {Weaver},
  {Zabalza}, {Edwards}, {Azalee Bostroem}, {Burke}, {Casey}, {Crawford},
  {Dencheva}, {Ely}, {Jenness}, {Labrie}, {Lim}, {Pierfederici}, {Pontzen},
  {Ptak}, {Refsdal}, {Servillat}, \& {Streicher}}]{Astropy:2013}
{Astropy Collaboration}, {Robitaille}, T.~P., {Tollerud}, E.~J., {et~al.} 2013,
  \aap, 558, A33

\bibitem[{Balbinot {et~al.}(2013)Balbinot, Santiago, da~Costa, Maia, Majewski,
  Nidever, Rocha-Pinto, Thomas, Wechsler, \& Yanny}]{Balbinot:2013}
Balbinot, E., Santiago, B.~X., da~Costa, L., {et~al.} 2013, \apj, 767, 101.
\newblock \url{http://stacks.iop.org/0004-637X/767/i=2/a=101}

\bibitem[{Bechtol {et~al.}(2015)Bechtol, Drlica-Wagner, Balbinot,
  {et~al.}}]{Bechtol:2015wya}
Bechtol, K., Drlica-Wagner, A., Balbinot, E., {et~al.} 2015, \apj, 807, 50

\bibitem[{{Belokurov} {et~al.}(2014){Belokurov}, {Irwin}, {Koposov}, {Evans},
  {Gonzalez-Solares}, {Metcalfe}, \& {Shanks}}]{Belokurov:2014}
{Belokurov}, V., {Irwin}, M.~J., {Koposov}, S.~E., {et~al.} 2014, \mnras, 441,
  2124

\bibitem[{{Bertin}(2006)}]{Bertin:2006}
{Bertin}, E. 2006, in Astronomical Society of the Pacific Conference Series,
  Vol. 351, Astronomical Data Analysis Software and Systems XV, ed.
  C.~{Gabriel}, C.~{Arviset}, D.~{Ponz}, \& S.~{Enrique}, 112

\bibitem[{{Bertin}(2011)}]{Bertin:2011}
{Bertin}, E. 2011, in Astronomical Society of the Pacific Conference Series,
  Vol. 442, Astronomical Data Analysis Software and Systems XX, ed. I.~N.
  {Evans}, A.~{Accomazzi}, D.~J. {Mink}, \& A.~H. {Rots}, 435

\bibitem[{{Bertin} \& {Arnouts}(1996)}]{Bertin:1996}
{Bertin}, E., \& {Arnouts}, S. 1996, \aaps, 117, 393

\bibitem[{Bianchini {et~al.}(2017)Bianchini, Sills, \&
  Miholics}]{Bianchini:2017}
Bianchini, P., Sills, A., \& Miholics, M. 2017, Monthly Notices of the Royal
  Astronomical Society, 471, 1181.
\newblock \url{https://dx.doi.org/10.1093/mnras/stx1680}

\bibitem[{{Bressan} {et~al.}(2012){Bressan}, {Marigo}, {Girardi}, {Salasnich},
  {Dal Cero}, {Rubele}, \& {Nanni}}]{Bressan:2012}
{Bressan}, A., {Marigo}, P., {Girardi}, L., {et~al.} 2012, \mnras, 427, 127

\bibitem[{{Bullock} \& {Johnston}(2005)}]{Bullock:2005}
{Bullock}, J.~S., \& {Johnston}, K.~V. 2005, \apj, 635, 931

\bibitem[{{Da Costa}(2003)}]{DaCosta:2003}
{Da Costa}, G.~S. 2003, in Astronomical Society of the Pacific Conference
  Series, Vol. 296, New Horizons in Globular Cluster Astronomy, ed.
  G.~{Piotto}, G.~{Meylan}, S.~G. {Djorgovski}, \& M.~{Riello}, 545

\bibitem[{{Da Costa} \& {Armandroff}(1995)}]{DaCosta:1995}
{Da Costa}, G.~S., \& {Armandroff}, T.~E. 1995, \aj, 109, 2533

\bibitem[{{Deason} {et~al.}(2015){Deason}, {Wetzel}, {Garrison-Kimmel}, \&
  {Belokurov}}]{Deason:2015}
{Deason}, A.~J., {Wetzel}, A.~R., {Garrison-Kimmel}, S., \& {Belokurov}, V.
  2015, \mnras, 453, 3568

\bibitem[{{DES Collaboration}(2018)}]{DES:2018}
{DES Collaboration}. 2018, \apjs, 239, 18

\bibitem[{{Dotter} {et~al.}(2010){Dotter}, {Sarajedini}, {Anderson},
  {Aparicio}, {Bedin}, {Chaboyer}, {Majewski}, {Mar{\'\i}n-Franch}, {Milone},
  {Paust}, {Piotto}, {Reid}, {Rosenberg}, \& {Siegel}}]{Dotter:2010}
{Dotter}, A., {Sarajedini}, A., {Anderson}, J., {et~al.} 2010, \apj, 708, 698

\bibitem[{{Drlica-Wagner} {et~al.}(2016){Drlica-Wagner}, {Bechtol}, {Allam},
  {et~al.}}]{Drlica-Wagner:2016}
{Drlica-Wagner}, A., {Bechtol}, K., {Allam}, S., {et~al.} 2016, \apjl, 833, L5

\bibitem[{Drlica-Wagner {et~al.}(2015)Drlica-Wagner, Bechtol, Rykoff,
  {et~al.}}]{Drlica-Wagner:2015ufc}
Drlica-Wagner, A., Bechtol, K., Rykoff, E.~S., {et~al.} 2015, \apj, 813, 109

\bibitem[{{Drlica-Wagner} {et~al.}(2018){Drlica-Wagner}, {Sevilla-Noarbe},
  {Rykoff}, {et~al.}}]{Drlica-Wagner:2018}
{Drlica-Wagner}, A., {Sevilla-Noarbe}, I., {Rykoff}, E.~S., {et~al.} 2018,
  \apjs, 235, 33

\bibitem[{{Eilers} {et~al.}(2019){Eilers}, {Hogg}, {Rix}, \&
  {Ness}}]{Eilers:2018}
{Eilers}, A.-C., {Hogg}, D.~W., {Rix}, H.-W., \& {Ness}, M. 2019, \apj, 871,
  120

\bibitem[{{Fadely} {et~al.}(2011){Fadely}, {Willman}, {Geha}, {Walsh},
  {Mu{\~n}oz}, {Jerjen}, {Vargas}, \& {Da Costa}}]{Fadely:2011}
{Fadely}, R., {Willman}, B., {Geha}, M., {et~al.} 2011, \aj, 142, 88

\bibitem[{Flaugher {et~al.}(2015)Flaugher, Diehl, Honscheid,
  {et~al.}}]{Flaugher:2015}
Flaugher, B., Diehl, H.~T., Honscheid, K., {et~al.} 2015, \aj, 150, 150

\bibitem[{Forbes \& Bridges(2010)}]{Forbes:2010}
Forbes, D.~A., \& Bridges, T. 2010, Monthly Notices of the Royal Astronomical
  Society, 404, 1203.
\newblock \url{http://dx.doi.org/10.1111/j.1365-2966.2010.16373.x}

\bibitem[{{Foreman-Mackey} {et~al.}(2013){Foreman-Mackey}, {Hogg}, {Lang}, \&
  {Goodman}}]{Foreman-Mackey:2013}
{Foreman-Mackey}, D., {Hogg}, D.~W., {Lang}, D., \& {Goodman}, J. 2013, \pasp,
  125, 306

\bibitem[{{Gaia Collaboration} {et~al.}(2018){Gaia Collaboration}, {Brown},
  {Vallenari}, {Prusti}, {de Bruijne}, {Babusiaux}, {Bailer-Jones}, {Biermann},
  {Evans}, {Eyer}, {Jansen}, {Jordi}, {Klioner}, {Lammers}, {Lindegren},
  {Luri}, {Mignard}, {Panem}, {Pourbaix}, {Randich}, {Sartoretti}, {Siddiqui},
  {Soubiran}, {van Leeuwen}, {Walton}, {Arenou}, {Bastian}, {Cropper},
  {Drimmel}, {Katz}, {Lattanzi}, {Bakker}, {Cacciari}, {Casta{\~n}eda},
  {Chaoul}, {Cheek}, {De Angeli}, {Fabricius}, {Guerra}, {Holl}, {Masana},
  {Messineo}, {Mowlavi}, {Nienartowicz}, {Panuzzo}, {Portell}, {Riello},
  {Seabroke}, {Tanga}, {Th{\'e}venin}, {Gracia-Abril}, {Comoretto},
  {Garcia-Reinaldos}, {Teyssier}, {Altmann}, {Andrae}, {Audard},
  {Bellas-Velidis}, {Benson}, {Berthier}, {Blomme}, {Burgess}, {Busso},
  {Carry}, {Cellino}, {Clementini}, {Clotet}, {Creevey}, {Davidson}, {De
  Ridder}, {Delchambre}, {Dell'Oro}, {Ducourant}, {Fern{\'a}ndez-
  Hern{\'a}ndez}, {Fouesneau}, {Fr{\'e}mat}, {Galluccio}, {Garc{\'\i}a-Torres},
  {Gonz{\'a}lez-N{\'u}{\~n}ez}, {Gonz{\'a}lez-Vidal}, {Gosset}, {Guy},
  {Halbwachs}, {Hambly}, {Harrison}, {Hern{\'a}ndez}, {Hestroffer}, {Hodgkin},
  {Hutton}, {Jasniewicz}, {Jean-Antoine-Piccolo}, {Jordan}, {Korn},
  {Krone-Martins}, {Lanzafame}, {Lebzelter}, {L{\"o}ffler}, {Manteiga},
  {Marrese}, {Mart{\'\i}n-Fleitas}, {Moitinho}, {Mora}, {Muinonen}, {Osinde},
  {Pancino}, {Pauwels}, {Petit}, {Recio-Blanco}, {Richards}, {Rimoldini},
  {Robin}, {Sarro}, {Siopis}, {Smith}, {Sozzetti}, {S{\"u}veges}, {Torra}, {van
  Reeven}, {Abbas}, {Abreu Aramburu}, {Accart}, {Aerts}, {Altavilla},
  {{\'A}lvarez}, {Alvarez}, {Alves}, {Anderson}, {Andrei}, {Anglada Varela},
  {Antiche}, {Antoja}, {Arcay}, {Astraatmadja}, {Bach}, {Baker},
  {Balaguer-N{\'u}{\~n}ez}, {Balm}, {Barache}, {Barata}, {Barbato}, {Barblan},
  {Barklem}, {Barrado}, {Barros}, {Barstow}, {Bartholom{\'e} Mu{\~n}oz},
  {Bassilana}, {Becciani}, {Bellazzini}, {Berihuete}, {Bertone}, {Bianchi},
  {Bienaym{\'e}}, {Blanco-Cuaresma}, {Boch}, {Boeche}, {Bombrun}, {Borrachero},
  {Bossini}, {Bouquillon}, {Bourda}, {Bragaglia}, {Bramante}, {Breddels},
  {Bressan}, {Brouillet}, {Br{\"u}semeister}, {Brugaletta}, {Bucciarelli},
  {Burlacu}, {Busonero}, {Butkevich}, {Buzzi}, {Caffau}, {Cancelliere},
  {Cannizzaro}, {Cantat-Gaudin}, {Carballo}, {Carlucci}, {Carrasco},
  {Casamiquela}, {Castellani}, {Castro-Ginard}, {Charlot}, {Chemin},
  {Chiavassa}, {Cocozza}, {Costigan}, {Cowell}, {Crifo}, {Crosta}, {Crowley},
  {Cuypers}, {Dafonte}, {Damerdji}, {Dapergolas}, {David}, {David}, {de
  Laverny}, {De Luise}, {De March}, {de Martino}, {de Souza}, {de Torres},
  {Debosscher}, {del Pozo}, {Delbo}, {Delgado}, {Delgado}, {Di Matteo},
  {Diakite}, {Diener}, {Distefano}, {Dolding}, {Drazinos}, {Dur{\'a}n},
  {Edvardsson}, {Enke}, {Eriksson}, {Esquej}, {Eynard Bontemps}, {Fabre},
  {Fabrizio}, {Faigler}, {Falc{\~a}o}, {Farr{\`a}s Casas}, {Federici},
  {Fedorets}, {Fernique}, {Figueras}, {Filippi}, {Findeisen}, {Fonti},
  {Fraile}, {Fraser}, {Fr{\'e}zouls}, {Gai}, {Galleti}, {Garabato},
  {Garc{\'\i}a-Sedano}, {Garofalo}, {Garralda}, {Gavel}, {Gavras}, {Gerssen},
  {Geyer}, {Giacobbe}, {Gilmore}, {Girona}, {Giuffrida}, {Glass}, {Gomes},
  {Granvik}, {Gueguen}, {Guerrier}, {Guiraud}, {Guti{\'e}rrez-S{\'a}nchez},
  {Haigron}, {Hatzidimitriou}, {Hauser}, {Haywood}, {Heiter}, {Helmi}, {Heu},
  {Hilger}, {Hobbs}, {Hofmann}, {Holland}, {Huckle}, {Hypki}, {Icardi},
  {Jan{\ss}en}, {Jevardat de Fombelle}, {Jonker}, {Juh{\'a}sz}, {Julbe},
  {Karampelas}, {Kewley}, {Klar}, {Kochoska}, {Kohley}, {Kolenberg},
  {Kontizas}, {Kontizas}, {Koposov}, {Kordopatis}, {Kostrzewa-Rutkowska},
  {Koubsky}, {Lambert}, {Lanza}, {Lasne}, {Lavigne}, {Le Fustec}, {Le
  Poncin-Lafitte}, {Lebreton}, {Leccia}, {Leclerc}, {Lecoeur-Taibi},
  {Lenhardt}, {Leroux}, {Liao}, {Licata}, {Lindstr{\o}m}, {Lister}, {Livanou},
  {Lobel}, {L{\'o}pez}, {Managau}, {Mann}, {Mantelet}, {Marchal}, {Marchant},
  {Marconi}, {Marinoni}, {Marschalk{\'o}}, {Marshall}, {Martino}, {Marton},
  {Mary}, {Massari}, {Matijevi{\v{c}}}, {Mazeh}, {McMillan}, {Messina},
  {Michalik}, {Millar}, {Molina}, {Molinaro}, {Moln{\'a}r}, {Montegriffo},
  {Mor}, {Morbidelli}, {Morel}, {Morris}, {Mulone}, {Muraveva}, {Musella},
  {Nelemans}, {Nicastro}, {Noval}, {O'Mullane}, {Ord{\'e}novic},
  {Ord{\'o}{\~n}ez-Blanco}, {Osborne}, {Pagani}, {Pagano}, {Pailler},
  {Palacin}, {Palaversa}, {Panahi}, {Pawlak}, {Piersimoni}, {Pineau}, {Plachy},
  {Plum}, {Poggio}, {Poujoulet}, {Pr{\v{s}}a}, {Pulone}, {Racero}, {Ragaini},
  {Rambaux}, {Ramos-Lerate}, {Regibo}, {Reyl{\'e}}, {Riclet}, {Ripepi}, {Riva},
  {Rivard}, {Rixon}, {Roegiers}, {Roelens}, {Romero-G{\'o}mez}, {Rowell},
  {Royer}, {Ruiz-Dern}, {Sadowski}, {Sagrist{\`a} Sell{\'e}s}, {Sahlmann},
  {Salgado}, {Salguero}, {Sanna}, {Santana- Ros}, {Sarasso}, {Savietto},
  {Schultheis}, {Sciacca}, {Segol}, {Segovia}, {S{\'e}gransan}, {Shih},
  {Siltala}, {Silva}, {Smart}, {Smith}, {Solano}, {Solitro}, {Sordo}, {Soria
  Nieto}, {Souchay}, {Spagna}, {Spoto}, {Stampa}, {Steele},
  {Steidelm{\"u}ller}, {Stephenson}, {Stoev}, {Suess}, {Surdej}, {Szabados},
  {Szegedi-Elek}, {Tapiador}, {Taris}, {Tauran}, {Taylor}, {Teixeira},
  {Terrett}, {Teyssandier}, {Thuillot}, {Titarenko}, {Torra Clotet}, {Turon},
  {Ulla}, {Utrilla}, {Uzzi}, {Vaillant}, {Valentini}, {Valette}, {van Elteren},
  {Van Hemelryck}, {van Leeuwen}, {Vaschetto}, {Vecchiato}, {Veljanoski},
  {Viala}, {Vicente}, {Vogt}, {von Essen}, {Voss}, {Votruba}, {Voutsinas},
  {Walmsley}, {Weiler}, {Wertz}, {Wevers}, {Wyrzykowski}, {Yoldas},
  {{\v{Z}}erjal}, {Ziaeepour}, {Zorec}, {Zschocke}, {Zucker}, {Zurbach}, \&
  {Zwitter}}]{Gaia:2018}
{Gaia Collaboration}, {Brown}, A.~G.~A., {Vallenari}, A., {et~al.} 2018, \aap,
  616, A1

\bibitem[{{Gilmore} {et~al.}(2007){Gilmore}, {Wilkinson}, {Wyse}, {Kleyna},
  {Koch}, {Evans}, \& {Grebel}}]{Gilmore:2007}
{Gilmore}, G., {Wilkinson}, M.~I., {Wyse}, R.~F.~G., {et~al.} 2007, \apj, 663,
  948

\bibitem[{{Gnedin} \& {Ostriker}(1997)}]{Gnedin:1997}
{Gnedin}, O.~Y., \& {Ostriker}, J.~P. 1997, \apj, 474, 223

\bibitem[{{G{\'o}rski} {et~al.}(2005){G{\'o}rski}, {Hivon}, {Banday},
  {Wandelt}, {Hansen}, {Reinecke}, \& {Bartelmann}}]{Gorski:2005}
{G{\'o}rski}, K.~M., {Hivon}, E., {Banday}, A.~J., {et~al.} 2005, \apj, 622,
  759

\bibitem[{{Harris}(1996)}]{Harris:1996}
{Harris}, W.~E. 1996, \aj, 112, 1487

\bibitem[{{Henden} \& {Munari}(2014)}]{Henden:2014}
{Henden}, A., \& {Munari}, U. 2014, Contributions of the Astronomical
  Observatory Skalnate Pleso, 43, 518

\bibitem[{Hunter(2007)}]{Hunter:2007}
Hunter, J.~D. 2007, Computing In Science \& Engineering, 9, 90

\bibitem[{{Jethwa} {et~al.}(2016){Jethwa}, {Erkal}, \&
  {Belokurov}}]{Jethwa:2016}
{Jethwa}, P., {Erkal}, D., \& {Belokurov}, V. 2016, \mnras, 461, 2212

\bibitem[{{Johnson} {et~al.}(1999){Johnson}, {Bolte}, {Stetson}, {Hesser}, \&
  {Somerville}}]{Johnson:1999}
{Johnson}, J.~A., {Bolte}, M., {Stetson}, P.~B., {Hesser}, J.~E., \&
  {Somerville}, R.~S. 1999, \apj, 527, 199

\bibitem[{Jones {et~al.}(2001)Jones, Oliphant, Peterson, {et~al.}}]{scipy:2001}
Jones, E., Oliphant, T., Peterson, P., {et~al.} 2001, {SciPy}: Open source
  scientific tools for {Python}, , .
\newblock \url{http://www.scipy.org/}

\bibitem[{{Kallivayalil} {et~al.}(2018){Kallivayalil}, {Sales}, {Zivick},
  {Fritz}, {Del Pino}, {Sohn}, {Besla}, {van der Marel}, {Navarro}, \&
  {Sacchi}}]{Kallivayalil:2018}
{Kallivayalil}, N., {Sales}, L., {Zivick}, P., {et~al.} 2018, \apj, 867, 19

\bibitem[{Keller {et~al.}(2011)Keller, Mackey, \& Costa}]{Keller:2011}
Keller, S.~C., Mackey, D., \& Costa, G. S.~D. 2011, The Astrophysical Journal,
  744, 57.
\newblock \url{https://doi.org/10.1088%2F0004-637x%2F744%2F1%2F57}

\bibitem[{{Kharchenko} {et~al.}(2013){Kharchenko}, {Piskunov}, {Schilbach},
  {et~al.}}]{Kharchenko:2013}
{Kharchenko}, N.~V., {Piskunov}, A.~E., {Schilbach}, E., {et~al.} 2013, \aap,
  558, A53

\bibitem[{{Kim} {et~al.}(2016){Kim}, {Jerjen}, {Mackey}, {Da Costa}, \&
  {Milone}}]{Kim:2016}
{Kim}, D., {Jerjen}, H., {Mackey}, D., {Da Costa}, G.~S., \& {Milone}, A.~P.
  2016, \apj, 820, 119

\bibitem[{{Kim} {et~al.}(2015){Kim}, {Jerjen}, {Milone}, {Mackey}, \& {Da
  Costa}}]{Kim:2015a}
{Kim}, D., {Jerjen}, H., {Milone}, A.~P., {Mackey}, D., \& {Da Costa}, G.~S.
  2015, \apj, 803, 63

\bibitem[{{Koposov} {et~al.}(2007){Koposov}, {de Jong}, {Belokurov}, {Rix},
  {Zucker}, {Evans}, {Gilmore}, {Irwin}, \& {Bell}}]{Koposov:2007}
{Koposov}, S., {de Jong}, J.~T.~A., {Belokurov}, V., {et~al.} 2007, \apj, 669,
  337

\bibitem[{{Koposov} {et~al.}(2017){Koposov}, {Belokurov}, \&
  {Torrealba}}]{Koposov:2017}
{Koposov}, S.~E., {Belokurov}, V., \& {Torrealba}, G. 2017, \mnras, 470, 2702

\bibitem[{{Koposov} {et~al.}(2015){Koposov}, {Belokurov}, {Torrealba}, \&
  {Evans}}]{Koposov:2015cua}
{Koposov}, S.~E., {Belokurov}, V., {Torrealba}, G., \& {Evans}, N.~W. 2015,
  \apj, 805, 130

\bibitem[{{Laevens} {et~al.}(2014){Laevens}, {Martin}, {Sesar}, {Bernard},
  {Rix}, {Slater}, {Bell}, {Ferguson}, {Schlafly}, {Burgett}, {Chambers},
  {Denneau}, {Draper}, {Kaiser}, {Kudritzki}, {Magnier}, {Metcalfe}, {Morgan},
  {Price}, {Sweeney}, {Tonry}, {Wainscoat}, \& {Waters}}]{Laevens:2014}
{Laevens}, B.~P.~M., {Martin}, N.~F., {Sesar}, B., {et~al.} 2014, \apjl, 786,
  L3

\bibitem[{{Laevens} {et~al.}(2015){Laevens}, {Martin}, {Bernard}, {Schlafly},
  {Sesar}, {Rix}, {Bell}, {Ferguson}, {Slater}, {Sweeney}, {Wyse}, {Huxor},
  {Burgett}, {Chambers}, {Draper}, {Hodapp}, {Kaiser}, {Magnier}, {Metcalfe},
  {Tonry}, {Wainscoat}, \& {Waters}}]{Laevens:2015b}
{Laevens}, B.~P.~M., {Martin}, N.~F., {Bernard}, E.~J., {et~al.} 2015, \apj,
  813, 44

\bibitem[{{Leaman} {et~al.}(2013){Leaman}, {VandenBerg}, \&
  {Mendel}}]{Leaman:2013}
{Leaman}, R., {VandenBerg}, D.~A., \& {Mendel}, J.~T. 2013, \mnras, 436, 122

\bibitem[{{Lindegren} {et~al.}(2018){Lindegren}, {Hern{\'a}ndez}, {Bombrun},
  {Klioner}, {Bastian}, {Ramos-Lerate}, {de Torres}, {Steidelm{\"u}ller},
  {Stephenson}, {Hobbs}, {Lammers}, {Biermann}, {Geyer}, {Hilger}, {Michalik},
  {Stampa}, {McMillan}, {Casta{\~n}eda}, {Clotet}, {Comoretto}, {Davidson},
  {Fabricius}, {Gracia}, {Hambly}, {Hutton}, {Mora}, {Portell}, {van Leeuwen},
  {Abbas}, {Abreu}, {Altmann}, {Andrei}, {Anglada}, {Balaguer-N{\'u}{\~n}ez},
  {Barache}, {Becciani}, {Bertone}, {Bianchi}, {Bouquillon}, {Bourda},
  {Br{\"u}semeister}, {Bucciarelli}, {Busonero}, {Buzzi}, {Cancelliere},
  {Carlucci}, {Charlot}, {Cheek}, {Crosta}, {Crowley}, {de Bruijne}, {de
  Felice}, {Drimmel}, {Esquej}, {Fienga}, {Fraile}, {Gai}, {Garralda},
  {Gonz{\'a}lez-Vidal}, {Guerra}, {Hauser}, {Hofmann}, {Holl}, {Jordan},
  {Lattanzi}, {Lenhardt}, {Liao}, {Licata}, {Lister}, {L{\"o}ffler},
  {Marchant}, {Martin-Fleitas}, {Messineo}, {Mignard}, {Morbidelli}, {Poggio},
  {Riva}, {Rowell}, {Salguero}, {Sarasso}, {Sciacca}, {Siddiqui}, {Smart},
  {Spagna}, {Steele}, {Taris}, {Torra}, {van Elteren}, {van Reeven}, \&
  {Vecchiato}}]{Lindegren:2018}
{Lindegren}, L., {Hern{\'a}ndez}, J., {Bombrun}, A., {et~al.} 2018, \aap, 616,
  A2

\bibitem[{{Luque} {et~al.}(2017){Luque}, {Pieres}, {Santiago}, {Yanny},
  {et~al.}}]{Luque:2017}
{Luque}, E., {Pieres}, A., {Santiago}, B., {Yanny}, B., {et~al.} 2017, \mnras,
  468, 97

\bibitem[{{Luque} {et~al.}(2016){Luque}, {Queiroz}, {Santiago}, {Pieres},
  {et~al.}}]{Luque:2016}
{Luque}, E., {Queiroz}, A., {Santiago}, B., {Pieres}, A., {et~al.} 2016,
  \mnras, 458, 603

\bibitem[{{Luque} {et~al.}(2018){Luque}, {Santiago}, {Pieres}, {Marshall},
  {et~al.}}]{Luque:2018}
{Luque}, E., {Santiago}, B., {Pieres}, A., {Marshall}, J.~L., {et~al.} 2018,
  \mnras, 478, 2006

\bibitem[{{Mackey} \& {Gilmore}(2004)}]{Mackey:2004}
{Mackey}, A.~D., \& {Gilmore}, G.~F. 2004, \mnras, 355, 504

\bibitem[{{Mackey} {et~al.}(2010){Mackey}, {Huxor}, {Ferguson}, {Irwin},
  {Tanvir}, {McConnachie}, {Ibata}, {Chapman}, \& {Lewis}}]{Mackey:2010}
{Mackey}, A.~D., {Huxor}, A.~P., {Ferguson}, A.~M.~N., {et~al.} 2010, \apj,
  717, L11

\bibitem[{{Mar{\'\i}n-Franch} {et~al.}(2009){Mar{\'\i}n-Franch}, {Aparicio},
  {Piotto}, {Rosenberg}, {Chaboyer}, {Sarajedini}, {Siegel}, {Anderson},
  {Bedin}, {Dotter}, {Hempel}, {King}, {Majewski}, {Milone}, {Paust}, \&
  {Reid}}]{Marin-Franch:2009}
{Mar{\'\i}n-Franch}, A., {Aparicio}, A., {Piotto}, G., {et~al.} 2009, \apj,
  694, 1498

\bibitem[{{Martin} {et~al.}(2008){Martin}, {de Jong}, \& {Rix}}]{Martin:2008}
{Martin}, N.~F., {de Jong}, J.~T.~A., \& {Rix}, H.-W. 2008, \apj, 684, 1075

\bibitem[{{Massari} {et~al.}(2017){Massari}, {Posti}, {Helmi}, {Fiorentino}, \&
  {Tolstoy}}]{Massari:2017}
{Massari}, D., {Posti}, L., {Helmi}, A., {Fiorentino}, G., \& {Tolstoy}, E.
  2017, \aap, 598, L9

\bibitem[{{McConnachie}(2012)}]{McConnachie:2012}
{McConnachie}, A.~W. 2012, \aj, 144, 4

\bibitem[{{Morganson} {et~al.}(2018){Morganson}, {Gruendl}, {Menanteau},
  {et~al.}}]{Morganson:2018}
{Morganson}, E., {Gruendl}, R.~A., {Menanteau}, F., {et~al.} 2018, \pasp, 130,
  074501

\bibitem[{{Mu{\~n}oz} {et~al.}(2012){Mu{\~n}oz}, {Geha}, {C{\^o}t{\'e}},
  {Vargas}, {Santana}, {Stetson}, {Simon}, \& {Djorgovski}}]{Munoz:2012}
{Mu{\~n}oz}, R.~R., {Geha}, M., {C{\^o}t{\'e}}, P., {et~al.} 2012, \apjl, 753,
  L15

\bibitem[{{Nidever} {et~al.}(2008){Nidever}, {Majewski}, \& {Butler
  Burton}}]{Nidever:2008}
{Nidever}, D.~L., {Majewski}, S.~R., \& {Butler Burton}, W. 2008, \apj, 679,
  432

\bibitem[{{Pace} \& {Li}(2019)}]{Pace:2018}
{Pace}, A.~B., \& {Li}, T.~S. 2019, \apj, 875, 77

\bibitem[{Pe\~{n}arrubia {et~al.}(2016)Pe\~{n}arrubia, G\'{o}mez, Besla, Erkal,
  \& Ma}]{Penarrubia:2016}
Pe\~{n}arrubia, J., G\'{o}mez, F.~A., Besla, G., Erkal, D., \& Ma, Y.-Z. 2016,
  Monthly Notices of the Royal Astronomical Society: Letters, 456, L54.
\newblock \url{http://dx.doi.org/10.1093/mnrasl/slv160}

\bibitem[{{Piskunov} {et~al.}(2007){Piskunov}, {Schilbach}, {Kharchenko},
  {R{\"o}ser}, \& {Scholz}}]{Piskunov:2007}
{Piskunov}, A.~E., {Schilbach}, E., {Kharchenko}, N.~V., {R{\"o}ser}, S., \&
  {Scholz}, R.-D. 2007, \aap, 468, 151

\bibitem[{{Schlegel} {et~al.}(1998){Schlegel}, {Finkbeiner}, \&
  {Davis}}]{Schlegel:1998}
{Schlegel}, D.~J., {Finkbeiner}, D.~P., \& {Davis}, M. 1998, \apj, 500, 525

\bibitem[{{Searle} \& {Zinn}(1978)}]{Searle:1978}
{Searle}, L., \& {Zinn}, R. 1978, \apj, 225, 357

\bibitem[{Skrutskie {et~al.}(2006)Skrutskie, Cutri, Stiening,
  {et~al.}}]{Skrutskie:2006}
Skrutskie, M.~F., Cutri, R.~M., Stiening, R., {et~al.} 2006, \aj, 131, 1163

\bibitem[{{Smith} {et~al.}(1998){Smith}, {Rich}, \& {Neill}}]{Smith:1998}
{Smith}, E.~O., {Rich}, R.~M., \& {Neill}, J.~D. 1998, \aj, 115, 2369

\bibitem[{{Torrealba} {et~al.}(2019){Torrealba}, {Belokurov}, \&
  {Koposov}}]{Torrealba:2019}
{Torrealba}, G., {Belokurov}, V., \& {Koposov}, S.~E. 2019, \mnras, 484, 2181

\bibitem[{{Torrealba} {et~al.}(2018){Torrealba}, {Belokurov}, {Koposov}, {Li},
  {Walker}, {Sanders}, {Geringer-Sameth}, {Zucker}, {Kuehn}, {Evans}, \&
  {Dehnen}}]{Torrealba:2018b}
{Torrealba}, G., {Belokurov}, V., {Koposov}, S.~E., {et~al.} 2018, arXiv
  e-prints, arXiv:1811.04082

\bibitem[{{Van Der Walt} {et~al.}(2011){Van Der Walt}, {Colbert}, \&
  {Varoquaux}}]{numpy:2011}
{Van Der Walt}, S., {Colbert}, S.~C., \& {Varoquaux}, G. 2011, Computing in
  Science \& Engineering, 13, 22

\bibitem[{{Watkins} {et~al.}(2019){Watkins}, {van der Marel}, {Sohn}, \&
  {Evans}}]{Watkins:2018}
{Watkins}, L.~L., {van der Marel}, R.~P., {Sohn}, S.~T., \& {Evans}, N.~W.
  2019, \apj, 873, 118

\bibitem[{{Wetzel} {et~al.}(2015){Wetzel}, {Deason}, \&
  {Garrison-Kimmel}}]{Wetzel:2015}
{Wetzel}, A.~R., {Deason}, A.~J., \& {Garrison-Kimmel}, S. 2015, \apj, 807, 49

\bibitem[{{Yozin} \& {Bekki}(2015)}]{Yozin:2015}
{Yozin}, C., \& {Bekki}, K. 2015, \mnras, 453, 2302

\bibitem[{{Zinn}(1993)}]{Zinn:1993a}
{Zinn}, R. 1993, in Astronomical Society of the Pacific Conference Series,
  Vol.~48, The Globular Cluster-Galaxy Connection, ed. G.~H. {Smith} \& J.~P.
  {Brodie}, 38

\end{thebibliography}

\end{document}